%% file: Main.tex
\documentclass[acmsmall,screen]{acmart}

\usepackage[bottom]{footmisc}
\usepackage{colortbl}
\usepackage[font=small,skip=1pt]{subcaption}
\usepackage{enumitem}
\usepackage{tikz}
\usepackage{cleveref}
\usepackage{multirow}
\usepackage{graphicx}
\usepackage{pdfpages}
\usepackage{array}

\definecolor{revisedtext}{RGB}{0, 0, 0}
\newcommand{\revised}[1]{\textcolor{revisedtext}{#1}}

\definecolor{revisedtextMin}{RGB}{0, 0, 0}
\newcommand{\revisedMin}[1]{\textcolor{revisedtextMin}{#1}}

\makeatletter
\newcommand\footnoteref[1]{\protected@xdef\@thefnmark{\ref{#1}}\@footnotemark}
\makeatother

\begin{document}

\title{``Can I Decorate My Teeth With Diamonds?'': Exploring Multi-Stakeholder Perspectives on Using VR to Reduce Children's Dental Anxiety}
\renewcommand{\shorttitle}{``Can I Decorate My Teeth With Diamonds?''}

\author{\href{https://orcid.org/0000-0002-9258-1169}{Yaxuan Mao}}
\email{stellayx@uw.edu}
\authornote{The author worked on this project while at City University of Hong Kong.}
\authornotemark[2]
\affiliation{
  \institution{University of Washington}
  \city{Seattle}
  \state{WA}
  \country{USA}
}

\author{\href{https://orcid.org/0000-0002-9767-3468}{Yanheng Li}}
\authornote{Equal contribution.}
\email{yanhengli3-c@my.cityu.edu.hk}
\affiliation{
  \institution{City University of Hong Kong}
  \country{Hong Kong SAR, China}
}

\author{\href{https://orcid.org/0009-0007-9098-1461}{Duo Gong}}
\email{12333216@mail.sustech.edu.cn}
\affiliation{
  \institution{Southern University of Science and Technology}
  \country{China}
}

\author{\href{https://orcid.org/0000-0002-7705-2031}{Pengcheng An}}
\email{anpc@sustech.edu.cn}
\affiliation{
  \institution{Southern University of Science and Technology}
  \country{China}
}

\author{\href{https://orcid.org/0000-0003-2016-4080}{Yuhan Luo}}
\email{yuhanluo@cityu.edu.hk}
\authornote{Corresponding author.}
\affiliation{
  \institution{City University of Hong Kong}
  \country{Hong Kong SAR, China}
}

\input{00-Abstract}

\begin{teaserfigure}
    \centering
    \includegraphics[width=\linewidth]{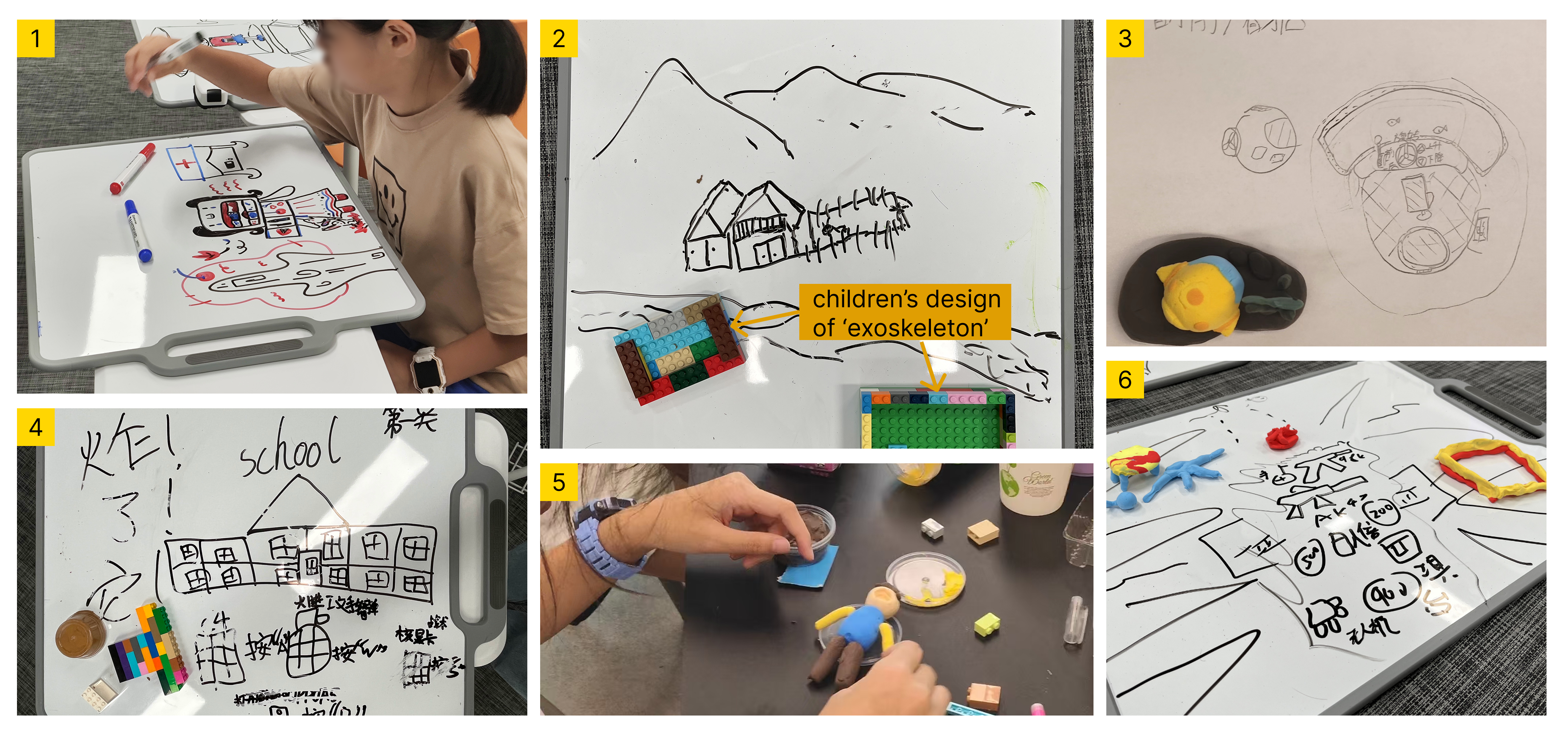}
    \caption{Designs \revised{created} by our children \revised{participants}: (1) \revised{transforming the dental clinic into a virtual} forest with adorable animals, \revised{where the child plays the role of a dentist treating a patient in a comforting environment}; (2) exploring a \revised{beautiful and calming} natural \revisedMin{environment} \revised{for relaxation, while engaging with the environment through a virtual exoskeleton interface}; (3) \revised{a virtual game like steel diver, which distincts from dental settings}; (4) \revised{simulating destruction of a school building for ``venting'' the discomfort and pain}; (5) \revised{role-playing as a dentist and learn} how to provide dental \revised{care to child} patients \revised{before the treatment; and} (6) \revised{a combat game} against zombies \revised{(metaphorical representation of tooth decay), which visualizes the treatment as a heroic and engaging challenge}.}
    \label{fig:designExample}
\end{teaserfigure}

\setcopyright{acmlicensed}
\acmJournal{PACMHCI}
\acmYear{2025} \acmVolume{9} \acmNumber{7} \acmArticle{CSCW346} \acmMonth{11}\acmDOI{10.1145/3757527}

\received{October 2024}
\received[revised]{April 2025}
\received[accepted]{August 2025}



\maketitle
\input{01-Introduction}
\input{02-RelatedWork}
\input{03-Co-designandInterviews}
\input{04-DataAnalysis}
\input{05-Results}
\input{06-Discussion}
\input{07-Conclusion}

\begin{acks}
We thank our participants for their interests and time, and anonymous reviewers for their helpful feedback. This project was supported by City University of Hong Kong (\#7005997 and \#7020106).
\end{acks}

\bibliographystyle{ACM-Reference-Format}
\bibliography{References}

\end{document}

%% file: 00-Abstract.tex
\begin{abstract}

Dental anxiety is prevalent among children, \revised{often leading} to missed treatment and potential negative effects on their mental well-being. 
While several interventions \revised{(e.g., pharmacological and psychotherapeutic} techniques) have been introduced \revised{for anxiety alleviation}, the \revised{recently emerged} virtual reality (VR) technology, with its immersive and playful nature, \revised{opened new opportunities for complementing and enhancing the therapeutic effects of existing interventions}. 
In this \revised{light}, we conducted \revised{a series of co-design} workshops with 13 children aged 10-12 to explore \revised{how they envisioned using VR to address their fear and stress associated with dental visits}, followed by interviews with parents (\textit{n} = 13) and two dentists. 
Our findings \revised{revealed} that children \revised{expected VR to provide} immediate relief, social support, and a sense of control \revised{during dental treatment}, parents \revised{sought} educational opportunities for \revised{their} children to \revised{learn about} oral health, and dentists prioritized treatment efficiency and safety issues.
Drawing from the findings, we discuss \revised{the considerations of multi-stakeholders for developing} VR-assisted anxiety management \revised{applications for children within and beyond dental settings.}

\end{abstract}

%% file: 01-Introduction.tex
\section{Introduction}

Dental anxiety refers to fear, anxiety, or stress associated with a dental setting, which not only manifests as fear of or resistance to dental treatment but is also associated with lasting emotional stress or discomfort~\cite{klingberg2007dental, seligman2017dental, gatchel1983prevalence}. This issue is particularly prevalent among children, as shown in prior research, about 26\% of children (six to 12 years old) worldwide have experienced dental anxiety~\cite{grisolia2021prevalence}. 
Compared \revised{to} adults, children have limited experience with pain and underdeveloped emotional regulation ability and are thus more vulnerable to \revised{the pain and discomforts during dental treatment}~\cite{Bhatia_2021}. As a result, \revised{children often} exhibit more serious resistance behaviors to dental visits, such as crying and tantrums, making the treatment procedure more challenging for themselves and the dentists~\cite{cohen2000impact}. In worse cases, such resistance can lead to missed treatment and reinforce negative associations with dental care~\cite{seligman2017dental, gatchel1983prevalence}.

\revised{Common} solutions to alleviate children's dental anxiety include \revised{both} pharmacological and psychotherapeutic interventions ~\cite{appukuttan2016strategies, malviya2000prolonged}.
Pharmacological interventions \revised{such as sedatives, are effective in managing physical pains and facilitating treatment, but do not necessarily address all factors contributing to dental anxiety, especially those related to mental stress and environmental influences~\cite{appukuttan2016strategies}. In worse cases, inappropriate use of medication can cause side effects on patients, such as motor imbalance and restlessness~\cite{malviya2000prolonged}.} 
Psychotherapeutic interventions, \revised{such as} cognitive behavior therapy (CBT), \revised{by helping individuals identify and change negative thoughts and behaviors, have gained} a higher acceptance~\cite{appukuttan2016strategies}. \revised{However, these interventions usually} require continuous involvement of professional therapists and specialized \revised{facility}~\cite{shahnavaz2016cognitive}, which are difficult to scale. 

In recent years, the emerging virtual reality (VR) technology has held great potential for supporting stress management and anxiety alleviation through immersive and playful experiences. Among existing research, VR has demonstrated its effectiveness in helping individuals overcome fear of needle~\cite{aydin2019using}, height~\cite{hodges1995virtual}, darkness~\cite{barcelo2020feasibility}, and being in water~\cite{montoya2024exploring}, and thus will likely to benefit children who face dental anxiety. 
However, existing work primarily \revised{targeted} adults, with only a few \revised{focusing on} children's anxiety \revised{alleviation, where VR is mainly used to play cartoon movies or relaxing scenes as a way to visually block the dental treatment~\cite{koticha2019effectiveness, niharika2018effects, du2022digital}. Although these studies found VR can help children reduce anxiety through quantitative measures, the rationale behind the intervention design remains unclear, for example, which components contributed to anxiety alleviation, and how VR functions beyond serving as a distraction technique. Besides, it is unclear how children subjectively think about and react to these interventions}.  

Furthermore, dental treatment typically involves \revised{communication and cooperation among} multiple stakeholders, including guardians (e.g., parents), dentists, and other clinic staff~\cite{shenkman2018stakeholder}. \revised{However, there has been limited research exploring their perspectives. For example, do children and their guardians hold similar or different views on using VR as a way to alleviate dental anxiety, and how does introducing these devices affect dentists' workflow? To bridge the gaps, we take these stakeholders' perspectives into consideration to examine their expectations of} using VR to alleviate children's dental anxiety, \revised{with the following research questions (RQs)}:



\begin{itemize}
  \item \textbf{RQ1.} Centering on \textbf{children's} needs, how do \textbf{children}, and \textbf{parents}, and \textbf{dentists} envision using VR to alleviate children's dental anxiety?
  \item \textbf{RQ2.} Centering on \textbf{parents'} and \textbf{dentists'} \revised{perspectives}, what are their \revised{attitudes and considerations} on using VR to alleviate children's dental anxiety?
\end{itemize}

\revised{Putting children at the center of design, we conducted a series of} co-design workshops with 13 children \revised{participants aged 10--12 who faced} dental anxiety, \revised{and interviewed} their guardians (parents) \revised{and} two dentists \revised{with children's design ideas presented as references}. 
\revised{In the co-design} workshops, each child participant talked about their recent dental visit and then interacted with three different VR applications. Next, \revised{they were encouraged to} design a VR-based scenario that they believed could help address their fear of dental \revised{treatment}. During this process, children were free to use various design widgets prepared by the research team while thinking aloud their design rationales. 

\revised{We found that} children were excited about VR technologies and created various design ideas: beyond using VR to distract \revised{them} from \revised{fearful} dental \revised{elements, they} envision an immersive experience that \revised{could} enhance their understanding of dental treatment procedures, connect \revised{them} to other peers with similar stress, and \revised{promote their} agency \revised{and sense of control} during the treatment. 
On the other hand, parents and dentists \revised{focused more on} leveraging VR for dental education, while raising concerns regarding safety implications \revised{and influences on the} clinical workflow. \revised{More importantly, they all} emphasized that \revised{VR interventions} cannot replace the role of parents' \revised{comfort and encouragement}. 


\revised{This study contributes} to the HCI \revised{and CSCW} community in \revisedMin{three folds}
: (1) an empirical understanding of the needs and challenges associated with alleviating dental anxiety in children, with considerations of multiple stakeholders' perspectives, including dental professionals, guardians, and the children themselves; (2) a synthesis of design considerations for developing virtual reality (VR) applications \revised{that address children's anxiety not only during dental treatment, but also opportunities to cultivate their awareness of oral health and self-regulation before, during, and after treatment}\revisedMin{; and (3) a thorough discussion on how these derived design elements can be applied to dental care in practice and extended to the broader context of anxiety alleviation, focusing on VR as a medium for empowerment, peer connection, and emotional regulation.} 

%% file: 02-RelatedWork.tex
\section{Related Work}

\subsection{Dental Anxiety and \revised{Coping} Strategies}

Dental anxiety encompasses feelings of fear, anxiety, or stress linked to dental environments, which manifests as apprehension or reluctance toward dental treatments~\cite{appukuttan2016strategies, klingberg2007dental, seligman2017dental}. \revised{When such anxiety persists, it can} contribute to ongoing emotional distress or \revised{even long-term mental} discomfort~\cite{gatchel1983prevalence, cohen2000impact}.
This condition is particularly prevalent among pediatric populations, as highlighted by meta-analyses of schoolchildren (aged between 4 and 12) with estimated occurrence rates of 27.6\%~\cite{grisolia2021prevalence}. Differing from other phobias like arachnophobia (fear of spiders) or acrophobia (fear of heights), which can be solved by keeping away from the fear source, dental anxiety is more challenging to manage because children who need dental treatment cannot simply avoid the treatment procedure. 
Among children, dental anxiety may delay or interrupt necessary \revised{and important} treatments, \revised{thereby} exacerbating their oral health problems, such as dental caries, tooth loss, and periodontal diseases~\cite{crego2014public}. 
These oral health issues \revised{can further cause more serious issues in children's health and growth~\cite{gilchrist2015impact}.}
Thus, \revised{effective management of} dental anxiety is \revised{crucial not only for dentists to complete the treatment but also} for the children to receive timely care.

Currently, \revised{dentists and other clinicians often combine} pharmacological and psychotherapeutic interventions \revised{to help children patients manage anxiety during treatment}~\cite{appukuttan2016strategies}. Pharmacological interventions \revised{refers to use of medication prior to the treatment, such as orally taking or inhaling sedative (e.g., nitrous oxide)~\cite{roelofse1996double} or use of psychoactive drugs (e.g., Benzodiazepines)} via intra-muscular, rectal, and intra-vascular administration routes~\cite{folayan2002review}. \revised{While reducing physiological discomfort, these approaches may not adequately eliminate anxiety related to other factors involving environmental, social, and mental stress~\cite{folayan2002review}.} 
Psychotherapeutic interventions \revised{such as cognitive behavioral therapy (CBT), focus on helping patients altering negative thoughts and behavioral patterns associated with dental visits, often through exposure, cognitive restructuring, and building positive experience~\cite{appukuttan2016strategies, armfield2013management}. Without potential side effects that might be introduced by pharmacological interventions~\cite{Kaar2016}, psychotherapeutic approaches are generally more acceptable. However, the scarcity and high cost of professional psychotherapists largely limit the scalability of these interventions~\cite{shahnavaz2016cognitive}.}

\revised{On the other hand, } virtual reality (VR) offers an alternative to these traditional coping strategies \revised{with the mobility of VR devices, and} provides immersive, playful experiences, \revised{holding promises to} reduce anxiety and discomfort \revised{in children on both physiological and mental levels, which we elaborate on below}.

\subsection{Virtual Reality (VR) \revised{Techniques in} Dental Anxiety Management}

\revised{There have been explorations of using VR to alleviate dental anxiety in the perioperative stage and during treatment operations~\cite{cunningham2021systematic, gujjar2019efficacy, Gujjar2017, lahti2020virtual}.
For example, Gujjar et al. applied} virtual reality exposure therapy (VRET) \revised{to help patients prepare for the treatment} through head-mounted display (HMD)~\cite{Gujjar2017, gujjar2019efficacy}. By exposing patients to 3D, stereoscopic dental environment, where they could navigate around to explore objects such as surgeon chair, lighting, and operation tools, \revised{the goal was to help them familiarize the operational environments and gradually overcome the fear of dental treatment in reality}~\cite{Gujjar2017}). 
\revised{Meanwhile, we have seen the use of} Virtual Reality Relaxation (VRR) strategies \revised{during the treatment procedure. In Lahti et al.'s work, patients were immersed in static} 360° videos of serene landscapes with enhanced audio and sound effects~\cite{lahti2020virtual}), \revised{which can visually and acoustically block the intense operation}, thereby mitigating \revised{their stress toward ongoing treatment}. 
Furthermore, some VR interventions \revised{are more interactive, such as by simulating movement such as walking or flying} through virtual botanical gardens~\cite{furman2009virtual}, or traverse beaches, forests, and mountains~\cite{wiederhold2014clinical}. 

\revised{While the above studies showed that} VR interventions can effectively reduce patients' stress and anxiety levels, they primarily focused on the adult population rather than children, \revised{whose reactions to anxiety-inducing situations differ significantly from adults due to underdeveloped cognitive and emotional functioning.}
\revised{For instance, children may experience fear more intensely and for longer periods compared to adults~\cite{kagan1983birth, carey1988conceptual}. Unlike adults, who can manage anxiety through intentional cognitive regulation, such as logical reasoning or breathe control, children} are more likely to exhibit stronger behavioral reactions, such as crying, clinging, or avoiding situations~\cite{mitchell2022developmental}. These differences suggest that interventions effective for adults, such as VRET, may not be equally effective for children and could even potentially exacerbate anxiety in the long term~\cite{matsuoka2017cognitive}. 

\revised{Among existing work, only a few utilized VR to alleviate dental anxiety for children, but they predominantly prioritized validating the effects of VR without articulating the rationales behind these intervention design~\cite{zaidman2023distraction, niharika2018effects, nunna2019comparative}.
For example, it is often briefly mentioned that children wore VR headsets playing cartoons or cinematic shows during treatment. Quantitative results, such as assessments of children's facial expressions, crying behaviors, and movements, indicated that VR helped reduce stress levels~\cite{nunna2019comparative, zaidman2023distraction, niharika2018effects}. However, these studies have not yet delved into how and why VR interventions work, or which parts of the experience contributed to children's reactions. Thus, there is still a lack of guidelines that designers can follow to effectively reduce dental anxiety in children. Besides, children's dental anxiety often stems from various sources beyond the ongoing treatment, including their memories of past unpleasant experiences, the ambiance of the dental environments (e.g., the presence of their parents), and even their guardians anxiety levels~\cite{kronicna2017psychosocial, brown1986psychological, dahlander2019factors}. The extent to which these factors are considered in the design of VR interventions remains unclear.}



\revised{Moreover, dental treatment is an inherent collaborative process among children, dentists, parents, and other clinical staff~\cite{yip2023co, jibb2023parent}. However, prior works rarely explored the perspectives of parents and dentists on VR as an anxiety alleviation technique. For instance, parents may concern about the safety, accessibility, or long-term effectiveness of VR interventions~\cite{yip2023co, jibb2023parent}, while dentists may be interested in how VR can be integrated into their workflow without disrupting treatment efficiency~\cite{o2023co, yip2023co, jibb2023parent}. To effectively make use of VR interventions in clinical practices, the perspectives of these stakeholders are thus important.}

%% file: 03-Co-designandInterviews.tex
\section{Methods}

\revised{To answer RQ1 (children's envisioned designs), we conduct co-design with children in groups. As shown in prior research, co-design is an effective approach to engage children and elicit their needs; specifically,  by asking children to visualize their design ideas, they are empowered to convey their thoughts through non-verbal cues, allowing researchers to cover insights that are often overlooked in pure interviews~\cite{van2016children, tsvyatkova2019review, read2002investigation}. 
In designing VR/AR technologies, \revised{``body storming'' is often used as a form of co-design,} allowing participants actively use their bodies and physical actions to simulate and explore design ideas in a 3D space~\cite{kitson2023co, o2023co, schaper2021co, Strickler_2019}. 
Taken together, we followed established frameworks to organize the co-design workshops by incorporating 
tangible prototyping materials including papers, pens, Lego kits, and plasticine~\cite{YourTurn, woodward2022would, yin2023children} to facilitate idea visualization; and body storming activities to explore the interactions in VR interactions~\cite{schaper2021co, Strickler_2019}.}

\revised{To complement RQ1 and answer RQ2 (parents' and dentists' needs), we employed semi-structured interviews with parents and dentists, while presenting children's designs as references. We did not involve them into the co-design sessions, as we aimed to create an environment where children participants could express their thoughts and ideas freely, without being influenced by adult expectations~\cite{van2016children}. Additionally, from a practical standpoint, separating these sessions saved coordination effort.} 
\revised{In this regard, our study consists of the following activities, which} were approved by the university's Institutional Review Board (IRB) after going through a full review process:

\begin{itemize}
    \item Five co-design sessions with children (\textit{n} = 13) to understand their challenges and needs in the face of dental anxiety, in which participants were asked to explore design ideas of using VR to alleviate such anxiety before, during, and after dental treatment.
    \item Parent interviews in the form of a group session based on their children's design ideas (\textit{n} = 13), which focused on parents' acceptance, expectations, and concerns about using VR to help alleviate children's dental anxiety.
    \item Dentist interviews based on children's and parents' ideas and expectations (\textit{n} = 2), which centered on how VR-based anxiety alleviation interventions may affect their workflow and treatment outcomes. 
\end{itemize}

\begin{figure}[h]
    \centering
    \includegraphics[width=\linewidth]{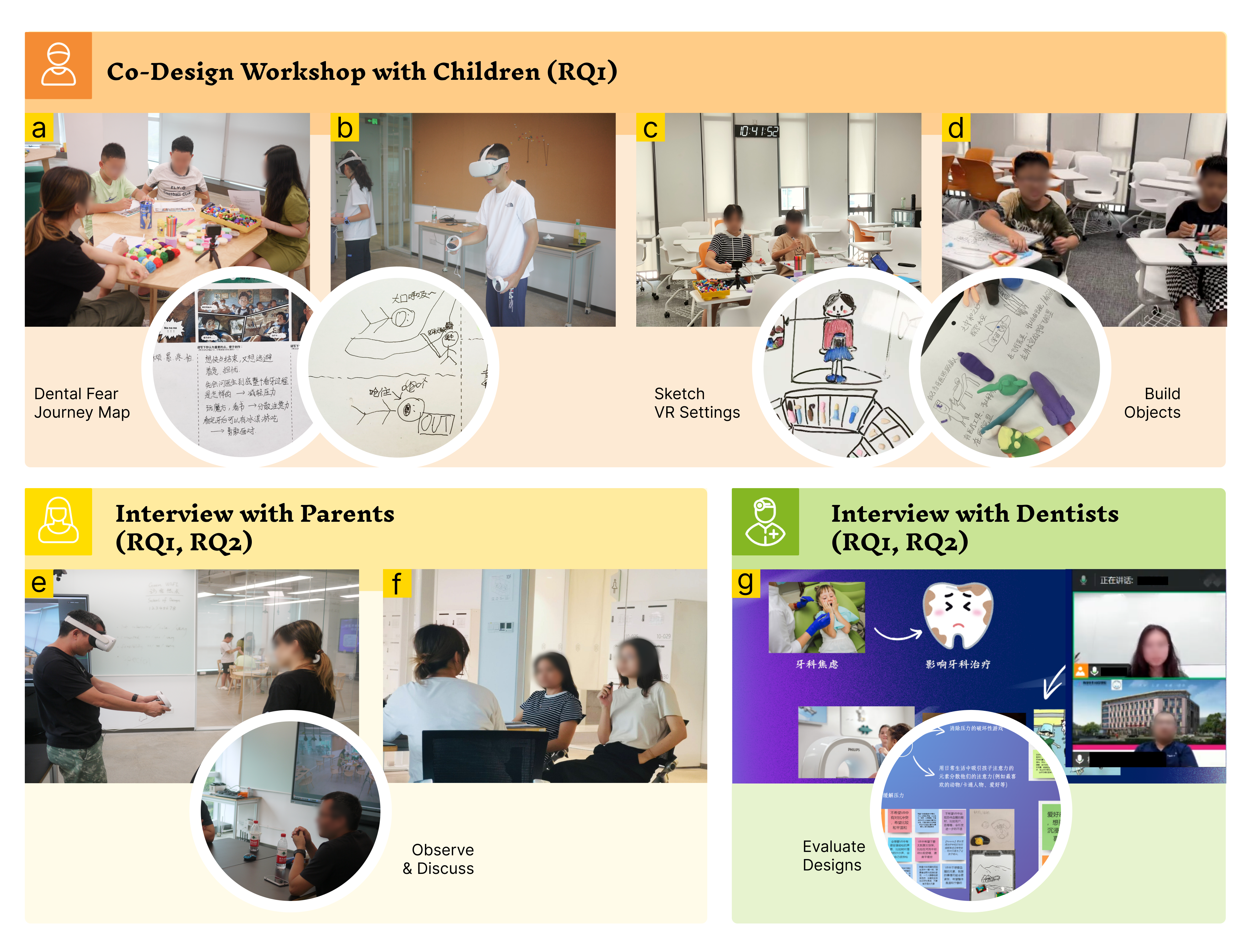}
    \caption{We Co-designed VR interventions to help children alleviate dental anxiety with three key stakeholders in three phases: co-design with children - warming up (a) \& VR inspiration session (b)\& design activity and discussion (c\&d); VR experience (e) and interviews (f) with guardians; interviews with dentists (g).}
    \label{fig:codesignProcess}
\end{figure}

\subsection{Participants}
\subsubsection{Children and guardians}
Our children and guardian participants were recruited from social media, posters distributed in the primary school, and snowball sampling. There were 44 guardians signed up for participation, and 13 met the following inclusion criteria: 
The child of the guardians (1) between 10 and 12 years old, (2) having received dental treatment within the past year, (3) without psychiatric/mental or developmental diagnoses (e.g., ADHD, autism, etc.), and (4) experiencing moderate or severe dental anxiety, assessed by the Children's Fear Survey Schedule---Dental Subscale (CFSS-DS)~\cite{klingberg1994dental, klingman1984effects, cuthbert1982screening}.
We targeted children aged 10 to 12 because Meta restricts access to Quest 2 and 3 for those under 10~\footnote{https://www.meta.com/quest/parent-info}. Additionally, dental anxiety is prevalent among children aged 7 to 12, whereas such anxiety is considerably less frequent in those above 12 years old~\cite{grisolia2021prevalence}. Given that children between 7 and 12 are largely in the same stage of cognitive development~\cite{fischer1985stages}, we focused on the 10 to 12 age group.

As presented in Table~\ref{table:01}, a total of 13 children, aged 10 to 12 (M = 11, SD = 1), were recruited for the workshop, comprising 6 girls and 7 boys. Given that the children are minors and to facilitate parallel interviews with parents during the workshops, the involvement of the child's guardian is essential for this study. Guardians were given the option to decide whether one or both parents would participate. As a result, we had 13 guardian participants in total. Specifically, the guardian of C1 was unavailable to participate, so C2's mother, who had a close relationship with C1's family, participated on behalf of C1's parents. For C6 and C8, both guardians took part. Since C10 and C11 are siblings (group 4), only one guardian was involved. For the remaining children, one guardian per child participated. At the end of the study, we provided $200$ RMB to each guardian participant.

\subsubsection{Dentists}
We also utilized a convenience sampling approach to recruit two dentists after the children's workshop, whose background information is shown in Table~\ref{table:02}. Both dentist participants had substantial expertise in treating pediatric patients within the age range of 10 to 12 years. They provided expert insights and were compensated at a rate of $100$ RMB each after dentists' interviews.


\begin{table*}[htbp]
\centering
\caption {\label{tab:table1} Demographic Information of Children and Guardians}
\resizebox{0.8\linewidth}{!}{
\begin{tabular}{ccccccc}
\hline
\textbf{Group}          & \textbf{Children ID} & \textbf{Gender} & \textbf{Age} & \textbf{CFSS-DS scores} & \textbf{Parent ID} & \textbf{Guardian gender} \\ \hline
\multirow{3}{*}{Group1} & C1                   & Male            & 12           & 64                      & N/A                & N/A                      \\
                        & C2                   & Female          & 12           & 55                      & P1                 & Female                   \\
                        & C3                   & Female          & 12           & 69                      & P2                 & Female                   \\ \hline
\multirow{3}{*}{Group2} & C4                   & Male            & 12           & 38                      & P3                 & Female                   \\
                        & C5                   & Female          & 10           & 39                      & P4                 & Female                   \\
                        & C6                   & Female          & 12           & 46                      & P5 and P6          & Female and Male          \\ \hline
\multirow{3}{*}{Group3} & C7                   & Male            & 10           & 64                      & P7                 & Female                   \\
                        & C8                   & Male            & 10           & 44                      & P8 and P9          & Female and Male          \\
                        & C9                   & Male            & 11           & 40                      & P10                & Male                     \\ \hline
\multirow{2}{*}{Group4} & C10                  & Female          & 12           & 45                      & P11                & Female                   \\
                        & C11                  & Female          & 10           & 48                      & P11                & Female                   \\ \hline
\multirow{2}{*}{Group5} & C12                  & Male            & 10           & 51                      & P12                & Male                     \\
                        & C13                  & Male            & 10           & 52                      & P13                & Male                     \\ \hline
\label{table:01}
\Description{Demographic Information of Dentists}
\end{tabular}
}
\end{table*}

\raggedbottom
\begin{table*}[htbp]
\centering
\caption {\label{tab:table1} Demographic Information of Dentists}
\resizebox{0.8\linewidth}{!}{
\begin{tabular} {lccccc}
\toprule
\textbf{Dentist ID}  &\textbf{Gender}   &\textbf{Age}         &\textbf{work seniority} &\textbf{Dental Specialties}    &\textbf{Experience of using VR} \\
\midrule
$D1$  & Male    & 45  & 24 years  & Pediatric Dentist   & Have used several times before   \\
$D2$  & Female  & 55  & 33 years  & General Dentist     & Only knowing the technology  \\
\bottomrule
\end{tabular}
}
\label{table:02}
\Description{Demographic Information of Dentists}
\end{table*}

\subsection{Co-Design Workshops with Children}
We conducted five co-design workshops involving a total of 13 children. Each session included 2-3 children. \revised{We chose this small group setting for two reasons: (1) compared to individual sessions, small-group co-design encourages interaction and collaboration among participants, and this dynamic often leads to a more engaging and supportive environment, where children can inspire one another and build on each other's ideas~\cite{fails2013methods, walsh2013facit}; and (2) compared to large group settings, each child in small groups has more opportunities to contribute, ensuring that all voices are heard, which balances communication quality and idea diversity~\cite{lowry2006impact}.} 

Upon arrival, children and their parents were presented with a consent form detailing the workshop activities and participants' rights. After a thorough review, parents provided their signatures, granting permission for audio and video recording during the workshops. \revised{We initially planned the co-design workshops to last about 150 minutes including a 10-minute break, which also aligned with previous co-design studies with children (one to three hours)~\cite{kitson2023co, dahl2020co, umulu2018wonder, lamonica2022developing}. 
However, in our practice, children participants spent more time on getting familiarized with VR interactions and were more active in creating design ideas than anticipated.  To ensure that they could restore and maintain focused~\cite{flavell1995development}, we arranged longer breaks (30 to 45 minutes). In the end, each workshop lasted 150 to 210 minutes.}

\subsubsection{Briefing and Warm-up}
Before the co-design activity, we provided children participants with a brief overview of the workshop's aim and procedure. Then, we asked each participant to recall their previous dental visits, how their parents prepared them for the visits, and how they felt before, during, and after the visits. This section \revised{was} designed to assist them in recalling their previous dental experiences, which \revised{helped contextualize their} subsequent design process. To \revised{make} participants better articulate their experiences, we encouraged them to describe or visualize their previous visits on a user journey map illustrated in Fig.~\ref{fig:codesignProcess}. Each stage in the map was accompanied by a narrative comic illustrating a potential scenario, their emotional state, and possible \revised{coping} strategies. 
In this briefing session, we \revised{asked children to} discuss: (1) key factors contributing to their dental anxiety at different stages; (2) strategies that children themselves or their parents \revised{and dentists} employed to help mitigate such anxiety; and (4) children's \revised{thoughts on} these strategies \revised{and how effective they were}. We also encouraged participants in the same session to share and discuss their experiences. The briefing session lasts approximately 20 minutes.

\subsubsection{VR Inspiration Session}
\begin{figure}
    \centering
    \includegraphics[width=\linewidth]{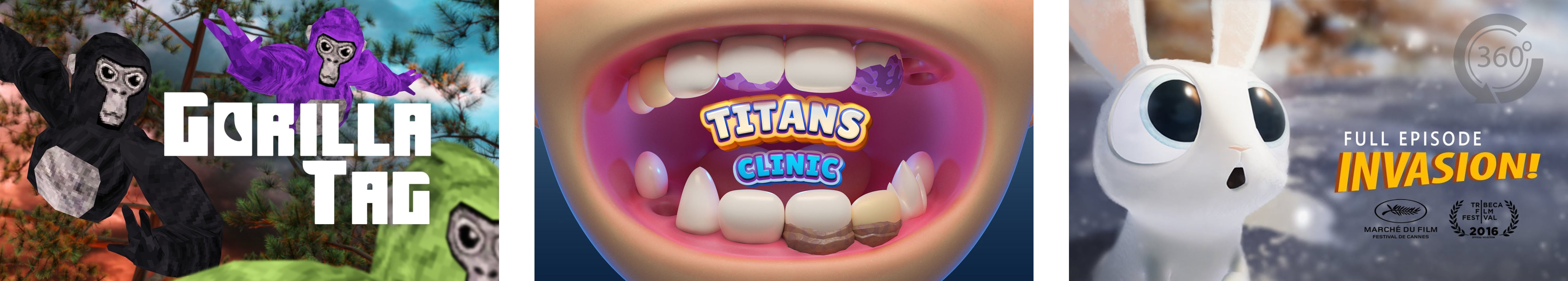}
    \caption{Three VR applications that our children participants interacted with during the study}
    \label{fig:VRApplications}
\end{figure}

To help children understand \revised{different types of interactions and} contents that can be implemented in a VR environment, we introduced three VR applications in Meta Quest 2~\footnote{https://www.meta.com/quest/products/quest-2/}, as shown in Fig.~\ref{fig:VRApplications}. 
\revised{These applications were selected to familiarize children with VR, ensuring that they understand the interaction norms so as to engage in the design activities followed. Our goal was to select distinctive VR applications that covered typical interaction mechanisms, while being appropriate for children aged 10 to 12. After a thorough search, we selected three applications: (1) Gorilla Tag\footnote{https://www.gorillatagvr.com}, an exploration game from the first-person perspective that incorporates extensive body movement and interaction in the VR space, where players become monkeys and navigate through a complex environment by physically moving their bodies, which introduces them to a highly interactive VR experience; (2) INVASION! Anniversary Edition~\footnote{https://www.youtube.com/watch?v=9hHeUqQbvVs}, a 360-degree film that showcases passive VR experiences, which allows viewers to experience a narrative in a fully immersive environment, providing a contrast experience to the interactive nature of the other applications; and (3) Titans Need Dental Care Too\footnote{https://www.meta.com/experiences/titans-clinic/6035422123217068/}, a simulated dental clinic designed to educate and familiarize children with dental procedures in a fun and engaging way, which serves as a practical example of how VR can be designed for dental care settings.}
\revised{Note that it is a common practice in co-design studies to showcase applications examples or give participants prepared widgets to help them quickly familiarize with the target design platform~\cite{luo2019co, qi2025participatory}, as the goal of co-design is to derive participants' needs behind their design rather than directly presenting their ideas as standard design guidelines~\cite{van2016children, tsvyatkova2019review, read2002investigation, qi2025participatory}.}


\revised{With two sets of Meta Quest 2, we gave each child 5 minutes to interact with each application.}
After interacting with the three VR applications, children were encouraged to share their experiences and discuss the possibilities of using VR to mitigate their dental anxiety. 
In particular, we asked each participant to refer to the user journey map they created during the briefing session and to think about when a VR-based intervention might be most helpful.

\subsubsection{Design Activity}
We structured the design activity into three steps. First, participants were instructed to focus on their challenges in managing dental anxiety, \revised{which they mentioned earlier}. \revised{In this step}, each participant worked individually to outline the settings and background for their own VR application on an A4 paper or a whiteboard~\cite{kelly2006bluebells, walsh2013facit}. If they were stuck or did not know where to start, one of the researchers would approach them and ask a few prompting questions, such as, ``what do you worry most during a dental visit'' and ``do you think any of your parents' or dentists' coping strategies can be translated to a VR environment?''
Second, \revised{we asked participants to} further \revised{elaborate on} their designs, such as \revised{by incorporating additional} characters and key props for interaction, or creating \revised{multimodal effects such as} sound that might help \revised{with anxiety alleviation} and enrich the user experience. 
Participants \revised{were free to use any of the design materials that we prepared} to construct \revised{their ``ideal dental care support.''} 
The third \revised{step} focused on designing interaction methods within the VR environment \revised{designed by participants}. \revised{This was a} body-storming \revised{activity}~\cite{schleicher2010bodystorming}, \revised{where participants subjected their own body movement to physically ideate the design solutions}.
\revised{In this process, participants imagined themselves within the VR environment and demonstrated how the interaction should occur through physical moves and gestures.}
On average, the design activity took about 90-120 minutes.

\subsubsection{Demonstration and Discussion}
Upon completing the design activities, children took a break and then presented their designs to researchers and their peers. 
The presentations \revised{included an overall descriptions of each child's} design and the potential interactions' \revised{mechanisms, along with explanations of} how the designs could assist in managing dental anxiety.
Throughout this process, researchers guided the children to reflect \revised{on their ideas with prompting} questions such as, ``Why did you design these elements?'' and ``Why do you think this design could help alleviate your dental anxiety?'' 
Following each child's presentation, other children participants were also encouraged to share their comments and thoughts on their peers' designs with questions such as, ``which aspects of this design do you appreciate'' or ``would you like to incorporate their design into yours, why or why not?'' This collaborative discussion aimed to \revised{help each participants rethink their designs for improvement}, while fostering additional creations. This session lasts approximately 30 minutes.

\subsection{Guardians' \revised{(Parents')} Interviews}
After signing the consent forms, guardians (parents) of the same groups were escorted to a separate room from their children. In this space, \revised{parents} were able to observe their children's design activities from an online meeting platform. This arrangement ensured that children would not be influenced by their parent's presence, which can mitigate potential pressure that inhibits children from freely expressing their thoughts.  
Two members of the research team conducted parent interviews parallel with the children's VR inspiration and after observing the children's design discussions. During these interviews, parents were invited to share their experiences of taking their children to dental clinics and their thoughts on the children's designs. \revised{The whole interview took around 90 minutes.}

\subsubsection{Interview on Dental Experience (Parallel With Children's \revised{VR Inspiration Session})}
In this stage, our interview questions primarily focused on: (1) children's feelings and behaviors associated with dental treatment; (2) their strategies to manage their children's anxiety; (3) \revised{parents} observed children using self-soothing methods; \revised{and} (4) the interventions \revised{ideas that parents had about alleviating} their children's dental anxiety. For each question, we asked the parents to provide specific examples to cross-validate children's responses and minimize misunderstandings between parents and children. This approach also helped the researchers quickly understand the children and facilitated a smoother transition for them to engage in the workshop activities.
Subsequently, parents were invited to interact with the same VR applications as their children did in the VR inspiration session. This was to provide them with an empathetic understanding of the VR environment. 

\subsubsection{Interview on Children's Designs (After Children 
\revised{Co-design Workshops})}
Following the children's co-design workshop, we invited parents to share their thoughts about their children's designs and their general expectations of using VR to alleviate children's dental anxiety, particularly regarding the content, visual style, safety, and privacy considerations.  
Drawing upon their previous experiences with VR, parents were also encouraged to share their design ideas, if any. This process aimed to complement children's designs from the perspectives of parents.

\subsection{Dentists' Interviews}
Given the busy schedule of dentists, interviews were conducted remotely. Each interview session lasted approximately one hour. 
Before the interview, we compiled and summarized the designs proposed by children and \revised{parents}, which were then presented sequentially to the dentist participants. 
Based on the designs from children and their parents' comments, we invited participating dentists to provide suggestions and express concerns.
This interview aimed to gain insights into dentists' clinical practice in managing children's dental anxiety, as well as the potential benefits and risks of deploying VR interventions in dental clinics.

%% file: 04-DataAnalysis.tex
\section{Data Analysis}
Our dataset includes video and audio recordings of children’s co-design workshops and the interviews with guardians and dentists, as well as the prototypes created by the children participants. All the audio recordings were transcribed to text. Three researchers individually analyzed the first two audio transcriptions, taking a bottom-up approach to generate initial codes, and \revised{regularly} met to \revised{ensure our initial codes reflected a consistent understanding of participants' needs and thoughts. Subsequently}, the remaining transcriptions were randomly distributed among three researchers for coding. Each researcher's codes were then checked by another researcher, who discussed any \revised{different interpretation of the same source data or ideas to elaborate on the codes} with the original coder. \revised{For instance, in one instance, C1 stated, ``\emph{If everyone is also afraid of going to the dentist, I wouldn’t be as anxious}''--one researcher believed that C1 expressed that socializing with others who shared similar experiences would help alleviate dental anxiety, while the other researcher interpreted it as C1 indicating that just knowing others are in the same situation would reduce anxiety. After discussion and reviewing relevant contextual information in the recording, we determined that the second researcher’s interpretation more accurately reflected C1's thoughts. Following this procedure}, the first author created an initial thematic map, which was iteratively revised through discussion with the project supervisor~\cite{braun2006using}. This process was repeated until a final set of themes emerged.
Note that an interrater reliability check was not performed for the analysis as we aimed to understand participants' needs and rationales behind the designs in this particular study rather than to generalize the design patterns~\cite{McDonald2019}. Instead, we ensured analysis reliability by independent coding and cross-checking among the three coders.

Despite the relatively small sample size, the study generated a rich corpus of transcribed data, comprising 9,878 Chinese characters from five co-design workshop recordings, 5,966 Chinese characters from five interview sessions with guardians, and 2,526 Chinese characters from interviews with two dentists. This dataset provides a detailed account of the children's creative expressions and needs, the perspectives of their caregivers, and the professional insights of dental experts. It enables us to uncover their diverse needs across multiple dimensions, informing the development of VR design strategies that are tailored to children with dental anxiety, family dynamics, and clinical requirements.

%% file: 05-Results.tex
\section{Results}
Our data gathered rich insights into the perspectives and needs of children, parents, and dentists in using VR to alleviate children’s dental anxiety across different dental visit stages. \revised{Before proceeding to answering our RQs, this section first describes the coping strategies that children, parents, and dentists currently use to manage children's dental anxiety. As descried in Table~\ref{tab:table3},}
\revised{these strategies included pharmacological interventions, reassurance, distraction techniques, dental education, and post-treatment rewards. While these strategies are widely used in practice, their effectiveness varies. For instance, children reported that reassurance and distraction techniques were less effective, while dental education and post-treatment rewards were more helpful, although did not fully eliminate their fear.}
\revised{Building upon these coping strategies, the following two sections answer} RQ1 (Children's envisioned designs) and RQ2 (Parents' and dentists' needs for support), respectively.

\begin{table*}[ht]
\centering
\color{revisedtext}
\caption {\label{tab:table3} Current coping strategies that participants described to manage children's dental anxiety. Participants' IDs were appended to quotes that reflected similar experiences.}
\resizebox{0.95\linewidth}{!}{
\begin{tabular}{m{2.5cm}|m{3cm}|m{9cm}}
\hline
\textbf{Strategy} & \textbf{Description} & \textbf{Examples} \\ \hline
\shortstack[l]{\textbf{Pharmacological} \\\textbf{approaches}, \\implemented by \\dentists}
& Anesthesia, oral and sedentary sedatives,  or nitrous oxide                    
& ``\emph{We typically use nitrous oxide or general anesthesia to alleviate dental anxiety in children. However, parents often express concerns about potential impacts on their child’s cognitive development}.'' (D2)
\\
\hline
\multirow{3}{2.5cm}{\textbf{Reassurance}, \shortstack[l]{typically \\implemented by \\dentists and \\parents}} 
& Verbal reassurance with comforting words and encouragement
& ``\emph{My mom will comfort me and say `Come on, you are very brave'}.'' (C10, C11, C13)
\newline``\emph{I'd like to tell my kids `Relax, everything will be fine'}.'' (P1, P11)\\ \cline{2-3}
& Physical reassurance with gentle contact  
& ``\emph{They (parents) will also touch me gently}.'' (C3, C11)
\newline``\emph{My mom/dad always hold my hand during the treatment}.'' (C3, C5, C7)\\ \cline{2-3}
& \shortstack[l]{Creating a relaxing \\clinic environment} 
& ``\emph{The walls of our clinic are painted with bright colors, most commonly light pink or light green, and decorated with cartoon animal stickers, such as rabbits and monkeys}.'' (D1)
\\
\hline
\multirow{3}{2.5cm}{\textbf{Distraction techniques}, used by dentists, parents, and children themselves}
& Children divert their attention by thinking about other matters  
& ``\emph{I like to keep my hands occupied during dental visits, such as solving a Rubik’s cube. Focusing on something unrelated to the procedure helps divert my attention and allows me to ignore the dentist’s actions}.'' (C1) \\ \cline{2-3}
& \shortstack[l]{Children can play \\games while waiting \\for the treatment}
& ``\emph{Before going to the dentist, I allow my child to play games that I usually don't permit, as a way to distract them}.'' (P4, P6, P10) \\ \cline{2-3}
& Visually block the treatment procedure
& ``\emph{The doctor covered my eyes with a blue-green paper}.'' (C7)
\newline ``\emph{Some treatment rooms have a display screen attached to the dental light, playing popular cartoons}.'' (P3, P4, P7, D2)
\\
\hline
\shortstack[l]{\textbf{Dental education}, \\provided by both \\dentists and parents}
& Educate children about dental procedures and oral health to improve their understanding of the treatment
& ``\emph{Before the dental visit, I help my child get mentally prepared by explaining what will happen. This reduces their fear of the unknown and alleviates anxiety}.'' (P1, P8, P11, P12)
\newline ``\emph{I use simple metaphors, like `the water spray is like an elephant's trunk') to explain the purpose of different dental tools, and let my patients experience some tools, such as feeling the water spray or air puff}.'' (D2)
\\
\hline
\shortstack[l]{\textbf{Post-treatment} \\\textbf{rewards}, \\given by parents \\and dentists}
& \shortstack[l]{Parents and dentists \\promise to give children \\rewards after \\completing dental \\treatment to motivate \\and comfort them}
& ``\emph{I am very afraid of getting a tooth pulled, but my parents promised to give me extra pocket money. Driven by the financial incentive, I am willing to cooperate}.'' (C7)
\newline ``\emph{The dentist keeps candies in the clinic, and the children become happy after receiving them. Every time they visit the dentist, they look forward to getting candies}.'' (P6, D1)
\\
\hline
\end{tabular}
}
\end{table*}

\subsection{RQ1: Children's VR Designs \& Parents' and Dentists' Thoughts}
Children's design \revised{ideas mostly} focused on the `before treatment' \revised{stage} \revised{to better prepare for the upcoming treatment} and `during treatment' \revised{stage} \revised{to either eliminate or reframe elements of the dental experience that they find fearful. These stages emerged from our bottom-up analysis of children's designs, where the `before treatment' stage typically refers to the time spent waiting at the clinic, and the `during treatment' stage when they are actively undergoing treatment}. 
\revised{Additionally, parents expanded on their children's based on their parenting experiences. Dentists, leveraging their expertise in pediatric dental care, provided insights on the use of VR for managing dental anxiety and shared thoughts on children's ideas from the practical perspectives.}
\revised{As such, while putting children's design ideas at the center, we also incorporated the considerations of parents and dentists.}

\subsubsection{Pre-Treatment Education and Preparation}~\label{pre-treatment}
\revised{First, we present designs of children participants on how they envisioned VR to help them learn about dental care procedures and get mentally prepared for the treatment, along with parents' and dentists' thoughts on incorporating educational materials and feasibility to implement similar approaches.}

\textbf{Facilitating learning through simulating the treatment environment.} Children participants expressed concerns about \revised{the uncertainty of what would happen in the dental clinic} (C1, C3, C6, C7, C8, C11) and the \revised{levels of physical pain they may experience} (C1, C5). To \revised{help themselves mentally} prepare for the upcoming visit, several participants designed a virtual clinic (C1, C3, C4, C10, C11), \revised{where they could interact with the equipment and instruments found} \revisedMin{in }a real clinic, such as dental chairs, toothbrushes, dental handpiece, mallet, and operatory lights, \revised{to become familiar with the treatment environment.} 
\revised{For example}, C3, C4, and C11 wanted to learn about \revised{different} dental instruments \revised{by virtually touching and using them, so as to} \revised{mitigate their fear of unknown}. C1 and C3 \revised{hoped to experience} each treatment step from the dentist's perspective, \revised{believing that this could prepare} them for what to expect \revised{in a more empowered way (as shown in Figure~\ref{fig:designExample} (5))}.

\revised{While children participants favored the virtual clinic idea, they held different views on its extent of authenticity.} \revised{C1 wanted the appearance and instrumental functionalities of virtual} dental clinic \revised{to} closely resemble an authentic one: ``\emph{it helps me to create a true sense of being in a real dental clinic while wearing the VR headset},'' 
but others expressed fear towards \revised{specific dental elements such as} sharp tools (C1) and bloody or unhealthy \revised{oral cavity} (C1, C5). In response, these participants suggested artistic alterations, like cartoonish styles to make the experience more acceptable (C1, C3, C5, C7). 
\revised{Regarding this, D1 noted} that cartoonization generally would not affect children's overall cognitive bias towards the treatment, \revised{but both D1 and D2 highlighted the importance for children to develop an objective view of the treatment environment and process without exaggerated illustrations of pain or giving them the illusion that the treatment is completely comfortable, which may} ``\emph{cause psychological discrepancies for children during actual treatment}.'' P1 shared the same viewpoint, believing that a realistic VR environment can be more instructive.

\begin{figure}
    \centering
    \includegraphics[width=\linewidth]{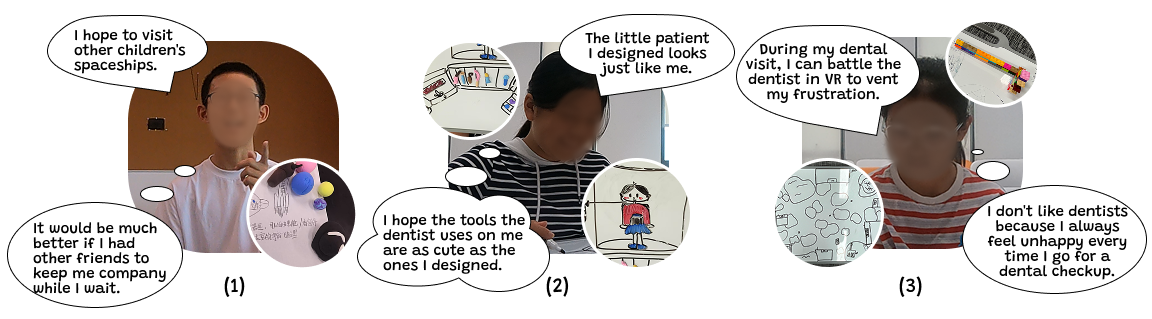}
    \caption{\revised{Children participants think-aloud during the design activity: (1) hoping to have peer accompany while waiting for treatment $\rightarrow$ designing a a virtual universe, where they can take spaceship to visit other children in the same situation; (2) being fear of sharp dental tools $\rightarrow$ making those tools appear in lovely shapes; and hoping to have nicely-looking teeth after treatment $\rightarrow$ role-playing as a dentist who gentaly treats a child patient look like themselves; (3) being fear of dentists $\rightarrow$ virtually fighting with the dentists in a game to vent out stress.}}
    \label{fig:ChildrenThinkChain}
\end{figure}

\textbf{Projecting \revised{positive treatment outcomes} via role-playing.}
\revised{With the uncertainty of the treatment outcomes}, four children wanted to play as a dentist in VR (C1, C4, C6, and C10), \revised{not only for better understanding the treatment procedure, but also} 
treating a virtual child patient resembles themselves. \revised{Through such role-playing activities, children hoped to} foster a sense of positive reciprocity, instilling \revised{the belief that} that they would receive similar care during their own dental treatments. C10, \revised{in particular, shared their thoughts of} decorating the teeth of the virtual patient, mirroring themselves, with attractive diamonds \revised{(see Figure~\ref{fig:ChildrenThinkChain} (2))}: 
``\emph{Can I decorate my teeth with diamonds? This small patient who looks like me is about to get braces and I think she would not like the brackets to be dark.}'' 

\revised{Similarly,} C1, C4, and C6 mentioned that \revised{as a ``professional little dentist,''} they would treat the child patient with utmost care and gentleness, offering comfort during moments of fear and focusing on restoring the their oral health. \revised{Behind these ideas, it was their hope for receiving} the same level of compassionate and attentive care. 
Parents and dentists \revised{also agreed that maintaining a positive emotion for the treatment is cruical to} ``\emph{facilitate communication between us and the child}.'' (D2).
D1 further \revised{commented on the role-playing design idea, sharing} that \revised{similar activities have been organized by some} private hospitals, where children can play as dentists, wear dental work clothes, and examine each other's teeth. \revised{Thus, D1 believed that} experiencing \revised{this activity in VR environment is} a fun and engaging way to alleviate children's anxiety that stemmed from the uncertainty of the dental treatment.


\textbf{Fostering peer support through shared \revised{dental} experiences.} 
Children participants expressed a desire to be connected with other peers who also faced dental anxiety through VR. \revised{They envisioned a multiplayer game where they could see and interact with other children who are also afraid of dental visits. This} shared \revised{experience} and struggles can give children a sense of \revised{assurance}, as C8 remarked, ``\emph{I hope it can support multiplayer mode so that I wouldn't feel lonely}'' \revised{(as shown in Figure~\ref{fig:ChildrenThinkChain} (1)).} \revised{Similarly, C1, C4, and C12 emphasized that being alone heightened their anxiety---\revised{while knowing that other children also face similar challenges would help them feel relieved and not ashamed about their anxiety. In this regard}, C7 and C9 designed an application to enable real-time interaction between different children via virtual avatars, voice, or text chat.}

\revised{In a similar vein}, C1, C5, and C11 wanted to interact with other pediatric patients who are also waiting for dental treatment in the VR environment, \revised{reassuring them} that they are not the only ones who feel scared. C12 said:
``\emph{If I can interact with other young patients who also need dental care in VR, I will feel that they are in the same situation as me. Mutual companionship can provide me with support, which is more useful than my mother's comfort. Because I feel that other children like me can better understand my feelings.}''

Parents participants P1, P7, and P11 \revised{recognized} that their children experienced reduced anxiety during dental procedures when peer-specific comfort strategies were adopted. \revised{P1 also helped children further explain,} ``\emph{Friends of the same age will have unique ways to comfort and be more able to relieve anxiety}''.
Dentist participants also concurred, acknowledging the potential of \revised{peer support}. Their \revised{viewpoint} is grounded in \revised{psychology of peer support}, which suggest that shared experiences and peer support can alleviate stress\revised{~\cite{richard2022scoping}}. However, current efforts to create play areas for peer socialization face practical \revised{challenges}:
``\emph{Public hospitals, constrained by space, struggle to provide areas for children to interact. While private hospitals may have more space, they often do not have many patients of similar age groups, making it difficult for children to build emotionally or socially meaningful relationships with their peers}.'' (D1)

\subsubsection{\revised{In}-treatment Engagement}~\label{in-treatment}
\revised{This subsection describes how children wanted to manage their fear and stress during the dental treatment, including ways to gain bodily agency and their varied preferences from serenity to tough-and-tumble play. They highlighted the importance to maintaining treatment awareness while reducing stressors, echoed by parents and dentists.}

\textbf{Creating \revised{bodily} agency.}
During dental treatments, children must remain still, and both C1 and C7 mention that the inability to move freely during such procedures introduces fear due to loss of control. Specifically, C7 noted that ``\emph{during dental treatment, having cotton stuffed in one's mouth makes speaking inconvenient and uncomfortable, evoking unpleasant associations with scenes from movies or TV shows where people are bound and gagged, leading to discomfort and insecurity.}'' Therefore, considering the therapeutic environment, children seek to regain a sense of control, which they cannot achieve in reality through VR.

\revised{In response,} children aim to attain bodily agency---the capacity to \revised{take} control \revised{of their body or the surroundings. This can be achieved either by hand-held} controllers \revised{to move and change characters, scenes, or music in VR or simulating movement within the virtual environment as if their own bodies are freely moving. For example,} C8 has designed an exoskeleton \revised{for himself to wear, which can help} manipulate numerous objects \revised{in the VR world} \revised{(see Figure~\ref{fig:designExample} (2)'s Lego)}, as stated by this \revised{participant}: 
``\emph{since I am lying on the dentist's chair unable to do anything, I would hope for an exoskeleton that could enable me to achieve many purposes through minimal hand movements alone.}'' 
\revised{Likewise,} C7 \revised{designed a similar control system}:
``\emph{I wish to have the VR control handle mounted on the chair in the clinic, allowing for a good control experience while preventing hand movements from interfering with the dentist's treatment.}''

Additionally, children \revised{chose to} enhance agency through proactive exploration \revised{of the} virtual world. Participants, including C1, C2, C5, C8, C9, and C11, envision vast, self-navigable virtual landscapes \revised{where they could virtually go anywhere}. C1 aspired to freely explore \revised{the universe} by piloting a spacecraft, C2 wishd to \revised{take} a submarine \revised{in the ocean, diving with} small fish and sunken ships (Figure~\ref{fig:designExample} (3)), C8 created an interactive rainforest \revised{to traverse through}, and C9 devised a space with weapon stations and medical rooms \revised{to conquer zombies and virtus} (Figure~\ref{fig:designExample} (6)). C11 \revised{highlighted the importance of such exploration freedom}:
``\emph{I hope for a diverse range of scenarios in VR, allowing for even more exploration. This way, I feel a sense of freedom, rather than being constrained or restricted.}''

\revised{Both} parents and dentists liked this idea, \revised{believing that granting bodily agency is a promising way for anxiety management}, as D1 added:
``\emph{Doctors certainly prefer patients to remain motionless in the dentist's chair. Therefore, addressing their anxiety stemming from a lack of physical control can be challenging. Utilizing VR to restore their agency is a highly effective approach.}''

\textbf{\revised{Varied preferences from serenity to rough-and-tumble play}.}
\revised{Our children participants expressed varied preferences for the VR environment---somefavored a calming and relaxing spaces  (two boys and five girls), while others wanted more exhilarating elements such as fighting games to vent their stress (five boys and one girl) .}
\revised{Those who preferred calming designs incorporated elements from their daily lives that they enjoyed}, such as their favorite animals (C11) and cartoon characters (C8). \revised{By creating an immersive world tailored to their interests, we can enhance the enjoyment of the experience and effectively sustain their attention through continuous distraction.} C11 expressed the wish to play with a little rabbit in VR while having their teeth checked, thus shifting their focus onto the rabbit. C5 agrees with this notion, stating that incorporating his favorite firearms and weapons into his design allows him to immerse himself in VR and forget his anxieties. Parents also highly endorse this approach; P2 and P10 believe that if the VR environment is filled with elements children love, it becomes much easier to distract them. Dentists concur with this view, as D1 added mentioned:
``\emph{changing the dental environment in real clinics is technically challenging, so using VR to immerse children in environments they enjoy can be very helpful.}''

Some children prefer destructive elements as a way to vent their emotions. Specifically, five boys and two girls suggested adding some functions that can be used for catharsis, including smash stones (C13), battle zombies (C5, C9), hack people (C2, C12), blow up schools (C7), and utilize violent tools to damage others' teeth (C4) \revised{(e.g., in Figure~\ref{fig:ChildrenThinkChain} (3), a children participant designed a VR application to fight with dentists)}.
C2 stated, ``\emph{I don't like dentists because I feel uncomfortable, so I want to kick them to vent my emotions.}'' C4 said, ``\emph{Destroying someone else's teeth in the virtual world doesn't make me anxious, because it doesn't matter to me what I destroy. I only want to vent my emotions.}'' C13 has similar feelings and desires to use various tools to strike rocks, hoping to obtain distinct and realistic tactile feedback when hitting rocks of different hardness levels. They explained:
``\emph{When I'm stressed, I strongly urge to vent my emotions. Breaking things is particularly satisfying because the tactile feedback I receive can make me feel very relieved.}''

Although dentists and parents expressed an understanding of children's need to decompress through destruction in VR, they imposed restrictions on the content. P2, P6, and D3 suggested that violent and gory elements should be avoided. P3, P7, P10, D1, and D2 all believed that the content should be positive and not contain negative influences. P7 remarked:
``\emph{Children should respect teachers and schools, and even if they want to vent, they should not blow up schools or harm teachers.}'' D1 expressed concern about the design of kicking dentists, fearing that such an outlet might foster antagonism towards dentists and worrying that children might confuse VR with reality, accidentally hurting a dentist in real life.

\textbf{\revised{Maintaining treatment awareness while reducing stressors}.}
Children often express anxiety triggered by various factors within the dental clinic, such as the sight of sharp instruments (C2, C6, C12, C13) \revised{and their} sounds (C1, C2, C4, C5, C7, C13), the \revised{seemly} smileless atmosphere created by dentists (C5, C6), and the clinical \revised{room} (C4, C5, C6, C11, C12). To mitigate \revised{such} stress, children \revised{wanted to be immersed in a more relaxing and peaceful} environment \revised{that} \revisedMin{is distinct}from the traditional dental clinic. Participants C2, C7, C9, C12, and C13 explicitly indicated a preference for VR experiences that are devoid of any dental-related elements during their treatment, seeking to evade the stressful clinical reality. 
This \revised{idea} resonated with both parents and dentists' \revised{views}, as P9 commented: 
``\emph{Maintaining children's entertainment during treatment, while avoiding elements that cause discomfort, is essential to reduce their fear and resistance.}''
Additionally, Some of the anxiety experienced by children stems from audio stressors. C7, C9, C10, and C11 all mentioned a fondness for calming background noises, such as the sound of running water, rustling leaves, and peaceful background music, as these auditory elements contributed to a relaxed atmosphere. 


\revised{While reducing the stressors, participants also shared that the unknowns during the operation procedure and their curiosity about their own oral health may introduce new stress (C4, C5, C6, and C9). }
\revised{As a result, these participants} proposed the integration of a real-time visual display \revised{of their mouth but with bloody spots or intense surgical operations such as cutting and drilling obscured}. Parents \revised{also} believed that exposing children to appropriate visuals of their oral conditions can help them become aware of the need to protect their teeth, as P6 noted that children over the age of 10 might be more tolerant of realistic visuals. \revised{Relatedly}, P7 proposed showing children the consequences of poor dental hygiene, such as the unappealing appearance of unhealthy teeth, to \revised{highlight} the importance of good dental care.


\subsection{RQ2: Parents' and Dentists' Needs and Concerns}
\revised{This section centers on parents and dentists' needs and concerns in addition to what they thought about children's designs.}
\revised{We noted that they both focused on long-term anxiety management rather than short-term alleviation, and brought up the possibility for children to receive `post-treatment' support after going through dental treatment for reflective learning.}

\subsubsection{Parents' Perspectives.}~\label{parents's perspectives}
Parents generally support their children's designs, \revised{but on the other hand,} they \revised{expect} that \revised{these} VR solutions not only address children's dental anxiety but also \revised{cultivate an} awareness of dental care. 
\revised{During the interviews, they also expressed} concerns regarding \revised{the health impact of VR interactions}.

\textbf{\revised{Fostering oral health awareness through VR education.}} 
\revised{Parents proposed two strategies to support this goal. This first focuses on reflective learning through post-treatment visualizations.} 
As P8 suggested, ``\emph{\revised{Adding a post-treatment VR visualization of the entire dental procedures may improve their comprehension and highlight the importance of oral hygiene habits.}}''
 \revised{The second emphasizes} integrating preventive oral care knowledge into daily VR experiences.
 \revised{For instance,} P10 proposed: ``\emph{Incorporating basic dental care knowledge into VR, particularly regarding the impact of daily diet and snack frequency on dental health \revised{, can enable children to implicitly learn how to maintain oral health, it can also prevent tooth decay by engaging them in virtual treatment}}.'' 
These suggestions \revised{reveal that parents see VR not only as a tool for anxiety reduction, but also as a means to educate children about oral health. By embedding preventive knowledge and reflective experiences in VR, they aim to instill long-term positive dental behaviors.}

\textbf{Concerns about health impacts of VR.} 
Almost all parents mentioned \revised{their concern about} the effect of prolonged VR use on children's eyesight. As P1 stated, ``\emph{\revised{I think about 30 minutes, like playing mobile games, should be okay. If it takes too long, that might hurt their eyes, but experts probably know better how to balance that while helping them stay calm.}}''
\revised{Some parents also worried about children being addicted to such immersive interactions}, particularly those involving games. ``\emph{We need a balanced approach in implementing VR for dental anxiety management, considering both its therapeutic benefits and potential health risks}.'' 

\subsubsection{Dentists' Perspectives.}~\label{dentists' perspectives}
\revised{Both two} dentists acknowledged that VR \revised{has potential to} help with children's dental anxiety. \revised{At the same time}, they also raised concerns \revised{regarding impact of using VR devices on clinical procedures and the ways to incorporate these devices with parents' involvement.}

\textbf{Virtual \revised{immersion} may \revised{interfere with physical treatment}.}
Dentists expressed concern about children becoming overly immersed in VR, potentially blurring the \revised{boundary} between virtual and physical reality. Given that most dental procedures require precision and necessitate patients to remain still and calm during treatment, dentists worried that exciting VR content or interactions requiring physical movement might be inappropriate to use during treatment. Instead, they suggested that soothing and relaxing designs would be more suitable to ensure treatment safety and accuracy. \revised{As D1 noted: ``\emph{Children can engage with VR applications that align with their interests and do not require physical movement, thereby reducing the risk of sudden motions that could pose medical hazards}.''}.

\textbf{VR as a \revised{supplement}, not a \revised{substitute for parents' support}.}
\revised{Dentists emphasized that} VR technology, \revised{or any other technolgy support}, cannot entirely replace \revised{existing} interventions. They \revised{noted that they might have overestimated the capability of} VR to \revised{serve as an ``omni-support'' system that covers} pre-treatment preparation, \revised{in-treatment} relaxation and engagement, and post-treatment education. \revised{On the one hand, the anxiety alleivation effects of VR may not be that powerful as expected; on the other hand,} such expectations may indicate an over-reliance on technology, potentially shifting responsibilities that parents themselves should bear onto technologies. D1 \revised{marked}: ``\emph{\revised{VR helps, but it doesn’t mean parents can step back. They still need to guide kids and support them throughout the entire process}}.''
\revised{Morever, dentists highlighted that children experiencing dental anxiety needs continuous support rather than one-time interventions. Thus, parents play a key role in shaping children's long-term attitudes toward dental care, while VR interaction can play a supplementary role.}

%% file: 06-Discussion.tex
\section{Discussion}
\revised{Our findings showed that all the stakeholders, including children, parents, and dentists hold positive views on using VR for dental anxiety alleviation. Unlike prior studies that merely took VR as a distraction tool, our study offers a more nuanced understanding of which VR elements can effectively address anxiety, how the visual and other sensory modalities should be designed to tailored individual preferences, and opportunities for VR interaction to enhance learning and awareness of oral health.
Meanwhile, parents and dentists exhibited concerns regarding the timing and usage of VR. Below, we discuss VR's unique benefits in engaging children and reduce their stress compared to existing strategies as well as important considerations for incorporating VR into dental care procedures.}

\subsection{VR's Unique \revised{Benefits in Alleviating Dental Anxiety} for Children}

\revised{Compared to existing solutions, VR offers a more immersive environment to engage children, keeping them staying away from the fear and stress, and can serve as an educational tool to help children learn about how to maintain their oral health. We elaborated on these unique aspects in the following.}

\subsubsection{\revised{Adaptive Virtual} Exposure.}
In section \revised{\ref{pre-treatment}}, children expressed a desire to familiarize themselves with the dental environment \revised{in a virtual} treatment setting, \revised{as a way to overcome} anxiety stemming from unfamiliarity (see \revised{Figure~\ref{fig:designExample} (5)}). \revised{These ideas} share \revised{a similar approach to} traditional exposure-based \revised{behavioral therapies}, \revised{which involve} as repeated exposure to real life stressors~\cite{appukuttan2016strategies}. \revised{Common strategies include techniques such as tell-show-do, a method involving verbal explanation of a procedure, visual and tactile demonstration, followed by actual performance~\cite{khandelwal2018control}, and modeling, where the child observes others, often the therapist, performing the desired behaviors~\cite{mowrer1966behavior}} 
\revised{However, some of our} children \revised{participants wanted} cartoonish representations of sharp \revised{dental} tools and bloody or unhealthy \revised{oral} parts, as realistic depictions could exacerbate their fears. \revised{Similar results were also reported in prior studies, suggesting} that prolonged exposure to \revised{realistic stressors} may inadvertently lead to increased fear and avoidance behaviors among children~\cite{bouton2002context, craske2006fear, schiller2010preventing, matsuoka2017cognitive}. 
\revised{This challenge highlights the potential of virtual exposure, which can be adapted to provide a less intimidating, more safe and personalized designs for anxiety management}. 


\revised{Drawing from our findings, VR designers can create exposure interactions specifically for children who are waiting for or in dental treatments. Specifically, the appearances of dental elements in the VR environment should be tailored to each child's interest and personality. As our children participants exhibited varied preferences in how to visualize dental clinic---regarding the visual styling and other sensory modalities including sound, tactile and even olfactory interactions~\cite{yan2023effectiveness, liu2019effect}. These applications}
could start with \revised{less realistic} simulations and gradually increase the \revised{authenticity} and intensity based on \revised{individual child user's learning} progress and comfort level~\cite{hinojo2020virtual}.
\revised{For instance, initial VR experiences might feature playful, cartoonish representations of dental tools and procedures, accompanied by calming background music or sound effects, such as soothing melodies or friendly voiceovers. As the child's anxiety diminishes and confidence grows, the VR environment can transition to more realistic representations, helping children become familiar with dental settings in a gradual and non-threatening manner.}
\revised{In certain cases, the authenticity of the virtual dental clinic can be blended with realistic backgrounds while selectively} blurring or obscuring visual stressors, as \revised{suggested by our children and dentist participants}. \revised{This approach enables children to experience a sense of realism without triggering unnecessary anxiety. Additionally, the inclusion of interactive elements, such as letting children virtually ``explore'' the clinic or play mini-games related to dental care, could further enhance engagement and reduce stress.} 

\subsubsection{\revisedMin{Sense of Control}.}
In sections \revised {\ref{pre-treatment} and \ref{in-treatment}, we observed that children mitigated dental anxiety not just by distracting themselves, but by actively role-playing positive treatment scenarios and immersing themselves in the virtual clinic} (see \revised{Figure~\ref{fig:designExample} (2) and (3)}).
\revisedMin{Unlike most existing interventions that primarily use VR to play cartoon movies or games, our findings} suggest that VR's potential goes beyond serving as a simple distraction tool. \revisedMin{VR can serve as a medium to provide} a sense of control or bodily agency, even if that control is not physical or direct. For instance, children in our study expressed a preference for interactive elements that allowed them to virtually move their bodies, simulating freedom from the physical restrictions of the dental chair. This included features such as using hand controllers to interact with game elements or navigate the virtual environment,  \revisedMin{which can help} alleviate the stress associated with being immobilized during treatment. This approach aligned with prior work that emphasized the role of control and agency~\cite{weems2003role}, where freedom of movement was found to can feelings of confinement and anxiety, fostering a more positive and engaging atmosphere. 
\revisedMin{By offering children an empowering way to feel more in control of their experience, an immersive VR intervention can also be applied to many other treatment scenarios where pediatric patients are restricted in movement.}

\subsubsection{\revised{Positive Projection Through Immersive} Learning.}
\revised{As reported in} section \revised{\ref{pre-treatment}}, children \revised{were eager learn about the procedure of} dental treatment (see \revised{Figure~\ref{fig:designExample} (1) and (5)}), \revised{which} parents and dentists \revised{resonate, as they all} believed that learning about dental knowledge and comprehending the treatment process can \revised{not only} alleviate \revised{dental anxiety but also encourage children to develop long-term habits to maintain their oral health}. This approach \revised{closely} aligns with traditional cognitive restructuring techniques, \revised{which aims at} alleviating anxiety by restructuring negative perceptions and replacing them with more accurate, positive cognition~\cite{de1995one}. \revised{In the context of dental anxiety, cognitive restructuring involves demystifying the treatment process and correcting misconceptions that often fuel fear, such as exaggerated concerns about pain or discomfort~\cite{appukuttan2016strategies, armfield2013management}. By offering children clear, age-appropriate explanations and educational experiences prior to treatment, this approach helps children feel more prepared.}


\revised{Within the VR environment,} the interactive nature \revised{can facilitate} cognitive restructuring \revised{in a} more \revised{immersive and} engaging \revised{manner}. By actively participating in the learning process, children \revised{are empowered to} challenge and reconstruct their negative \revised{perceptions} about dental treatments. For example, through role-playing as a dentist, \revised{they} can gain a deeper understanding of the treatment process from \revised{the dentists'} perspective, \revised{transforming a source of fear into a relatable and manageable experience}.
Furthermore, VR can enhance the learning experience \revised{by integrating} gamification elements \revised{to} create a fun and motivating educational environment, \revised{where children can receive} rewards \revised{such as achievement badges, positive feedback, and visible progress markers as they learn more about dental care knowledege}~\cite{nand2019engaging, nah2014gamification}. 
\revised{For example, children might earn points or unlock new levels as they ``treat'' virtual patients or learn about oral hygiene practices, which help sustain engagement, encourage repeated interaction, and make the learning process enjoyable}.


\subsection{Considerations for VR\revised{-based Anxiety Management in Children Within and Beyond Dental Care}}

While VR offers \revised{several benefits} in alleviating children's dental anxiety, interviews with dentists highlighted two key challenges in deploying such interventions. First, how to ensure treatment safety while children are immersed in VR? 
Second,  \revised{as} dental treatment for children is a team effort, successful outcomes depend on the active involvement of multiple stakeholders; \revised{thus, it is crucial to think about} how to facilitate effective collaboration among dentists, parents, and children to maximize the technology’s benefits.
\revised{Combing} our findings and prior literature, we elaborate on these challenges and potential research directions.

\subsubsection{Safeguarding During the Treatment}
\revised{The shared perspective between dentists and parents highlights a key challenge in using VR in pediatric dental settings: while immersion can effectively alleviate children's anxiety, it also poses risks to the treatment process} (Section \revised{\ref{parents's perspectives} and \ref{dentists' perspectives}}). Dentists emphasized the need for children to remain still and calm during treatments, where \revised{the operation needs to be} \revisedMin{precise in term of} \revised{ timing and oral parts}. They \revised{pointed out} that highly immersive or \revised{overly} interactive VR \revised{environment} might \revised{inadvertently lead to} physical movement, \revised{which can} potentially \revised{disrupt the treatment flow or even compromise procedural} safety. Some parents and children \revised{also expressed similar} concerns, \revised{emphasizing} the importance of \revised{of avoiding interactions that are too intensive or stimulating. While similar concerns about balancing immersion with practical care have been documented in other pediatric healthcare settings where VR is used for anxiety management~\cite{gold2021doc, lange2006virtual, aydin2019using, liszio2017virtual}, limited research has specifically addressed safeguarding during dental treatments, underscoring the need for more focused studies in this area}.

\revised{For the future work}, one \revised{direction} is to \revised{examine ways of prioritizing soothing and passive experiences that engage children without prompting physical movements or strong emotional reactions.}
\revised{Another direction involves the strategic allocation of VR usage across different stages of the dental treatment process.} One viable approach could be to dedicate more time to VR experiences during the pre-treatment stage. During this \revised{stage}, children can engage in \revised{more}immersive and interactive VR activities that help \revised{them vent stress and frustration. During the treatment, however, VR interventions should shift toward being more calming and lightweight, focusing on passive experiences that soothe children without encouraging movement or distraction}.



\subsubsection{Negotiation \revised{of} VR Use for \revised{Pediatric} Care}
Dentists emphasized the crucial role parents play in helping children manage dental anxiety, noting that VR should complement rather than replace parental involvement. 
While VR provides a novel avenue for home-based preparation, it also introduces potential tensions between children and parents regarding content choices.
Research on family media use suggests that digital technologies often require negotiation between parents and children, particularly regarding what content is appropriate~\cite{blackwell2016managing}. In the context of VR for \revised{anxiety management in the dental care context}, children may favor certain immersive experiences that parents deem unsuitable or ineffective. Effective negotiation between parents and children is essential to ensure that the chosen VR content adequately prepares children for dental visits while also aligning with parental expectations. This tension highlights the importance of respecting family dynamics as valuable cultural assets for learning and anxiety management~\cite{yu2023family}. 
To navigate \revised{this tension}, parents need to remain actively involved in guiding their children through VR-based preparation, ensuring the selected content aligns with both the child's preferences and their goals. \revised{At the same time}, these tensions could create opportunities for enhanced family discussions about dental knowledge. VR has the potential to spark conversations, promoting a better \revised{collective} understanding of dental health among family members.
Future VR design \revised{may} consider incorporating customizable features that allow parents and children to collaboratively select appropriate content, fostering a balance between entertainment and preparation for dental procedures.

Moreover, the adoption of VR in dental clinics and hospitals necessitates negotiation \revised{among a wide range of stakeholders, including not only children, parents, and dentists but also hospital staff, administrators, and technical teams responsible for implementation}.
Firstly, as dentists expressed in the interview, ``\emph{There are many considerations for hospitals to purchase new equipment.}'' \revised{Thus,} ensuring that the VR equipment is compatible with existing medical devices and systems \revised{is a key} to facilitate seamless integration into the healthcare workflow. This may involve technical assessments and partnerships with VR providers to ensure interoperability~\cite{gui2019making}. \revised{Hospitals should allocate sufficient resources} for the ongoing maintenance, updates, and troubleshooting of VR equipment~\cite{gui2019making}. 
\revised{Secondly,} another \revised{potential} area of negotiation involves integrating VR with patients' electronic health records (EHRs) and the sharing of \revised{patients' interaction} data with clinicians \revised{as references of their the treatment process and post-treatment reflection. This was noted during our} interviews \revised{with} parents, \revised{who} suggested utilizing VR to document the child's treatment process, serving as a visual reminder of their experiences. This recording process can be employed not only during treatment but also at home, capturing the child's condition and emotional fluctuations \revised{which} is important for maintaining continuity of care and enabling healthcare professionals to track the effectiveness of \revised{the interventions}~\cite{cowie2017electronic, tang2015restructuring}. 

\subsection{Limitations \revised{and Future Work}}

Despite our efforts to recruit participants from a broad demographic, the diversity within our sample (e.g., age, socioeconomic status, cultural background) is not fully representative. This study was conducted within a predominantly Chinese context, which may introduce cultural biases affecting generalizability to other populations. 
\revised{Relatedly, the sample size of 13 children and their guardians may be considered relatively small. This limitation stemmed from a practical constraint, as only 13 out of 44 children met the eligibility criteria (i.e, experiencing moderate or severe dental anxiety). However, we adopted an in-depth analysis approach, carefully examining each participant's design artifacts and activity transcripts. Given the our focus on multi-stakeholder perspectives, our study involved children, parents, and dentists, contributing to a rich understanding of their needs toward children's dental anxiety. Additionally, the sample size included in our work is similar to those reported in existing co-design studies with children~\cite{druin2002role, sanders2008co}.}

There may also be a potential selection bias, as children and their parents who agreed to participate might have had a pre-existing interest in technology or greater comfort with VR, making them more receptive to the intervention than the general pediatric population. The proposed VR system was conceptual rather than fully developed and tested. While children provided a detailed design, the lack of a functional prototype limits their ability to assess the practical effectiveness and usability of the system in real-world settings. This constraint may have hindered the full exploration and development of the idea, as actual implementation often reveals unforeseen challenges and opportunities for refinement that cannot be fully captured in theoretical prototypes.


\revised{Going forward, we plan to conduct field studies to further examine children, parents, and dentists' reactions to different designs of VR interventions, as well as the long-term effects of these designs.} 

%% file: 07-Conclusion.tex
\section{Conclusion}

In this \revised{work}, we report findings from co-design workshops involving 13 children aged 10--12, as well as interviews with their guardians and two dentists, to gain insights into how VR can be utilized to alleviate dental anxiety. During these workshops, the children shared their experiences with dental treatments and designed VR applications they believed would help ease their anxiety. Guardians and dentists, on the other hand, discussed their current methods for assisting children in managing dental anxiety, along with their considerations and concerns regarding the children's designs. Our \revised{findings revealed nuanced empirical insights into the perspectives of various stakeholders, including shared viewpoints and tensions regarding VR use.}
\revised{These insights contribute to the HCI and CSCW community by advancing understanding of how technology can be co-designed and implemented to address pediatric dental anxiety while accommodating the needs of children, parents, and clinical professionals like dentists}. 
